# Template engineering of Co-doped $BaFe_2As_2$ single-crystal thin films

S. Lee[1], J. Jiang[2], C. T. Nelson[3], C. W. Bark[1], J. D. Weiss[2], C. Tarantini[2], H. W. Jang[1], C. M. Folkman[1], S. H. Baek[1], A. Polyanskii[2], D. Abraimov[2], A. Yamamoto[2], Y. Zhang[3], X. Q. Pan[3], E. E. Hellstrom[2], D. C. Larbalestier[2], C. B. Eom[1*]

[1]Department of Materials Science and Engineering, University of Wisconsin-Madison, Madison, WI 53706, USA

[2]Applied Superconductivity Center, National High Magnetic Field Laboratory, Florida State University, 2031 East Paul Dirac Drive, Tallahassee, FL 32310, USA

[3]Department of Materials Science and Engineering, The University of Michigan, Ann Arbor, Michigan 48109, USA

**Understanding new superconductors requires high-quality epitaxial thin films to explore intrinsic electromagnetic properties, control grain boundaries and strain effects, and evaluate device applications[1-9]. So far superconducting properties of ferropnictide thin films appear compromised by imperfect epitaxial growth and poor connectivity of the superconducting phase[10-14]. Here we report novel template engineering using single-crystal intermediate layers of (001) $SrTiO_3$ and $BaTiO_3$ grown on various perovskite substrates that enables genuine epitaxial films of Co-doped $BaFe_2As_2$ with high transition temperature ($T_c$, $_{\rho=0}$ of 21.5K), small transition widths ($\Delta T_c$ = 1.3K), superior $J_c$ of 4.5 MA/cm$^2$ (4.2K, self field) and strong c-axis flux pinning. Implementing $SrTiO_3$ or $BaTiO_3$ templates to match the alkaline earth layer in the Ba-122 with the alkaline earth-oxygen layer in the templates opens new avenues for epitaxial growth of ferropnictides on multi-**

**functional single crystal substrates. Beyond superconductors, it provides a framework for growing heteroepitaxial intermetallic compounds on various substrates by matching interfacial layers between templates and thin film overlayers.**

Epitaxial pnictide thin films have so far been hard to produce, especially the F-doped highest $T_c$ rare earth (RE) 1111 phase[13-14]. Because both As and F are volatile at the deposition temperature, it is difficult to control the overall stoichiometry of the deposited film[13]. By contrast, Co, which can be used as the dopant in Ba(Fe,Co)$_2$As$_2$ (Ba122), has a low vapour pressure under growth conditions. Alkaline earth (AE) 122 phases have been grown on (001) oriented (La,Sr)(Al,Ta)O$_3$ (LSAT) and LaAlO$_3$ (LAO) single-crystal substrates by us and other groups[10-12]. However, the quality of the films reported so far is not satisfactory because $T_{c,\rho=0}$ and $\Delta T_c$ are 14~17K and 2~4K, respectively, values which are much lower and broader than those of bulk single crystals[15]. Furthermore, $J_c$ (5K, SF) of these films is ~10kA/cm$^2$, one to two orders of magnitude lower than in bulk single crystals[15]. We believe a fundamental reason for the low quality of the AE-122 phase on these substrates is that AE-122 is a metallic system, which does not bond well with oxide single-crystal substrates. In particular, LSAT and LAO substrates contain trivalent cations, whereas the alkaline earths of Ba-122 or Sr-122 are divalent, making the bonding between substrate and Ba-122 poor, which leads to a non-epitaxial, granular and poorly connected superconducting phase whose grains are separated by high-angle grain boundaries, wetting GB phases such as FeAs, and/or off-stoichiometric grains[16].

To overcome these problems, we explored the use of a template consisting of thin epitaxial layers of the divalent, AE-containing SrTiO$_3$ (STO) or BaTiO$_3$ (BTO) between various single-crystal perovskite substrates and the Ba-122 film. Building upon this common feature, we established the bonding model described in Fig.1. Recently, we reported a proof-of-principle of this concept by growing Co-doped Ba-122 epitaxial thin films on (001) oriented STO bicrystal substrates with intragrain $J_c$ as high as ~3 MA/cm$^2$ (4.2K, SF), values significantly higher than in previously reported epitaxial thin films[17]. However, the STO substrates became electrically conducting during deposition of the Co-doped Ba-122 thin film at 730 $^o$C in the high vacuum (2.7 × 10$^{-5}$ Pa), due to formation of oxygen vacancies in the STO[18]. This provides a parallel, current-sharing path between Ba-122 and STO which compromises normal-state property studies. Clearly normal-state behaviour and potential device applications[19] need insulating substrates, especially microwave devices which need substrates such as LSAT or LAO that have a low dielectric constant.

In this letter, we report single-crystal Co-doped Ba-122 epitaxial thin films using thin STO or BTO templates deposited on various perovskite single-crystal substrates, which include (001) LSAT, (001) LAO, and (110) GdScO$_3$, all of which yield Ba-122 films with superior superconducting properties on insulating substrates. The 50-100 unit cell (u.c.) thick STO and BTO templates were grown by pulsed-laser deposition (PLD) using a KrF excimer laser (248 nm) with high oxygen pressure reflection high energy electron diffraction (RHEED) for digital control and *in situ* monitoring of the epitaxial growth[20]. Co-doped Ba-122 films ~350 nm thick were grown on both bare single-crystal substrates and substrates with the STO and BTO templates. Structural and superconducting properties of these films are listed in Table SI in Supplementary information. We discuss here only films grown on bare STO, bare LSAT, STO/LSAT (i.e. an STO template on LSAT), and BTO/LSAT, while emphasizing that similar properties were obtained on (100) LAO and (110) GdScO$_3$ too.

Figure 2a and its inset show an atomic force microscope (AFM) image of a 100 u.c. STO template on LSAT and its RHEED pattern. They show atomically-flat terraces and single unit-cell high (~4 Å) steps, which confirm that the STO template layer is as good as those of bulk single crystal substrates of STO. It is found that an STO template up to 100 u.c. is fully coherent with the LSAT substrate. The epitaxial and crystalline quality of the Co-doped Ba-122 thin films were measured by four-circle x-ray diffraction (XRD). Figure 2b shows the out-of-plane $\theta$-$2\theta$ scans of the films on bare LSAT and 100 u.c. STO/LSAT. The XRD patterns show that the film 00$l$ reflections dominate, which indicates c-axis growth normal to the template and substrate. An extra peak at 2θ = ~65° in the film grown on STO/LSAT is the 002 reflection from Fe[11], however with intensity less than 0.5% of the Ba-122 (004) reflection. The intensity of the Ba-122 00$l$ peaks on STO/LSAT is about 2 orders of magnitude higher than that on bare LSAT. Rocking curves for the 004 reflection were measured to determine the out-of-plane mosaic spread and crystalline quality. As shown in Fig. 2c, the full width at half maximum (FWHM) of the 004 reflection rocking curve of the film on STO/LSAT is as narrow as 0.55°, which is the narrowest ever reported for AE-122 thin films, whereas that of the film on bare LSAT is as broad as 3.1°. Remarkably, the FWHM of the film on BTO/LSAT is as narrow as 0.17° similar to that of a Ba-122 bulk single crystal (Table SI)[21].

In-plane texture and epitaxial quality were determined by azimuthal $\phi$ scans of the off-axis (112) peak, as shown in Fig. 2d. The film grown on bare LSAT shows broad major peaks ($\Delta\phi$ =4.4°) every 90° with additional broader intermediate-angle peaks, which indicates that the in-plane Ba-122 structure consists of grains with high-angle tilt grain boundaries (GB). In contrast, the film grown on 100 u.c. STO/LSAT shows only sharp, strong peaks ($\Delta\phi$ =0.8°) every 90° characteristic of a truly epitaxial film with perfect in-plane texture.

To verify the crystalline quality, microstructures were studied by transmission electron microscopy (TEM). Figures 3a-c show the cross-sectional TEM images and selected area electron diffraction (SAED) patterns of the films grown on bare LSAT (Fig. 3a), 100 u.c. STO/LSAT (Fig. 3b), and bare STO (Fig. 3c). The films on bare STO or on 100 u.c. STO/LSAT show epitaxial relationships between the films and the substrate, while the film on bare LSAT shows a granular microstructure with many misoriented grains. The epitaxial films grown on bare STO and on STO/LSAT show line defects oriented along the c-axis as shown in Figs. 3b-c.

To characterize the superconducting transition temperature $T_c$, resistivity was measured as a function of temperature ($\rho$-T) by the van der Pauw method. As shown in Fig. 4a, the residual resistivity ratio (RRR) of the Ba-122 film grown on bare STO is as high as 160, a value consistent with oxygen deficient STO[18] that we reproduced on bare STO substrates heated in vacuum under the same conditions used for Ba-122 deposition. In contrast, the RRR of films grown on STO/LSAT and BTO/LSAT templates were 1.8 and 2.4, respectively, values characteristic of Co-doped Ba-122 single-crystals[22]. Additionally we noted that the room temperature resistivity of a film on bare LSAT is much higher than a film grown on templated LSAT, which we interpret as being due to strong scattering at the high-angle GBs. As shown in Fig. 4b, all films show high $T_{c,\,\rho=0}$ and narrow $\Delta T_c$ except for the film with high-angle GBs grown on bare LSAT. In particular, $T_{c,\rho=0}$ of the film on STO/LSAT is as high as 21.5K and $\Delta T_c$ is as narrow as 1.3K, which are the highest and narrowest values ever reported for AE-122 thin films.

Figure 4c shows the zero-field-cooled magnetization $T_c$ transitions. All three epitaxial films have a strong diamagnetic signal, in contrast to the polycrystalline film grown on bare LSAT, which, as shown by the inset in Fig. 4c, shows a three orders of magnitude smaller diamagnetic signal than the epitaxial films. This is a clear indication of magnetic granularity associated with the inability of screening currents to develop across the high-angle grain boundaries in the film on bare LSAT

Figure 5a shows $J_c$ as a function of magnetic field for all films determined from vibrating sample magnetometer measurements in fields up to 14 T. Here too all films except that on bare LSAT show high $J_c$, over 1 MA/cm$^2$ (4.2K, SF), which are the highest values ever reported for AE-122 thin films and are even better than in bulk single crystals (0.4 MA/cm$^2$ at 4.2K)[15]. Remarkably, the $J_c$ of the film on BTO/LSAT is as high as ~4.5 MA/cm$^2$ and the $J_c$ of the two epitaxial films furthermore has only a weak field dependence, indicative of little or no suppression of the $J_c$ by strong fields as indicated by the STO/LSAT film which had $J_c$ = ~0.4 MA/cm$^2$ even at 14 T. The magneto-optical (MO) image of the film on STO/LSAT (inset in Fig. 5a) shows strong flux shielding and a $J_c$ of ~3 MA/cm$^2$ (6.6K, 20 mT), similar to that deduced from the magnetization measurements. Such MO images confirm that the epitaxial films grown on STO/LSAT or BTO/LSAT are uniform and well-connected without any weak links.

We believe that two factors contribute to the high $J_c$. First, the films on templated LSAT and bare STO have high epitaxial quality with no high-angle tilt GBs, as confirmed by both XRD and TEM analysis. According to our previous report[17], [001] tilt grain boundaries of Ba-122 with θ = 6-24° show significant suppression of supercurrent, making it entirely understandable that randomly oriented high-angle tilt GBs would effectively block supercurrent, as suggested by several recent pnictide film reports[10-12,14]. Indeed the films grown on bare LSAT developed many high-angle GBs, had almost no flux shielding even in low fields (2 mT) and very low $J_c$. By contrast, films on templated LSAT and bare STO showed high $J_c$ without evidence of weak links. Interestingly the $J_c$ of the film on STO/LSAT is higher than that of the film on bare STO. We also measured the angular dependent transport $J_c$ of a STO/LSAT film, as shown in Fig. 5b. $J_c$ always show a strong c-axis peak, which opposes the expected electronic anisotropy since the upper critical field $H_{c2}$ is lower for H parallel to c-axis, making it clear that there is strong pinning along the Ba-122 c-axis that is parallel to the vertically aligned defects.

We have also grown Co-doped Ba-122 films on other bare substrates, including (001) LAO and (110) GdScO$_3$ (GSO). In spite of the almost perfect lattice match between Ba-122 and (110) oriented GSO, the film on bare GSO shows poor quality, just like the films grown on bare LAO (see Table SI). Both bare GSO and LAO films show broad peaks from misoriented grains in the $\phi$ scan. However, when an intermediate layer of STO or BTO was used as a template on the LAO or GSO, the properties of the Ba-122 films were dramatically enhanced and misoriented grains were not seen. Remarkably, we could also grow a superior quality Ba-122 film on STO/GSO using template engineering with intermediate STO layers which produced a perfect lattice match with Ba-122. The FWHM of the (004) reflection rocking curve and $\Delta\phi$ of Ba-122 on 50 u.c. STO/GSO were very narrow, 0.24° and 0.3°, respectively (See Table SI), values close to those of bulk single crystals[21]. All the above results show that an STO or BTO template is the key to growing high-quality epitaxial Ba-122 films.

In summary, we have developed a novel template engineering technique to turn granular, low-$J_c$ superconducting films into single-crystal, high-$J_c$, and truly epitaxial films of Ba-122. Jc in our high-quality epitaxial films is about 10 times greater than in bulk single crystals[15] and ~400 times greater than in previously reported AE-122 films[10-12]. In addition the crystallinity of our films is almost the same as bulk single crystals (Table S1). We believe that this template technique can be applied not only to perovskite single-crystal substrates, but also to other types of oxide substrates or even semiconductor substrates. Indeed, we have demonstrated this approach yields high-quality Co-doped Ba-122 films on STO-template grown on Si substrates [23-24] with $T_{c, \rho=0}$ = 18 K, $\Delta T_c$ = 1K, and no misoriented grains. This approach greatly expands substrate choices for high-quality Ba-122 thin films, thus allowing much broader fundamental property investigations of the newly discovered ferropnictide superconductors and parent compounds, as well as for exploring their applications. Furthermore, we expect epitaxial thin films of other layered intermetallics could be successfully grown on various types of oxide substrates by employing similar template engineering principles.

**Methods**

Co-doped Ba-122 thin films were grown *in-situ* on (001) oriented various single-crystal substrate and STO- or BTO- templated substrates using pulsed laser deposition (PLD) with a KrF (248 nm) UV excimer laser in a vacuum of $3 \times 10^{-4}$ Pa at 730 ~750 $^{\circ}$C. The base pressure before deposition was $3 \times 10^{-5}$ Pa, and the deposition took place at $3 \times 10^{-4}$ Pa due to degassing of the substrate heater. The Co-doped Ba-122 target was prepared by solid-state reaction with the nominal composition of Ba : Fe : Co : As = 1 : 1.84 : 0.16 : 2.2. The magnetization of the films with a size of about 2 mm x 4 mm was measured in a 14 T Oxford vibrating sample magnetometer (VSM) at 4.2K with the applied field perpendicular to the film surface. Magneto-optical imaging was used to examine the uniformity of current flow in the films so as to validate the use of the Bean model for converting the magnetic moment measured in the VSM to $J_c$ assuming current circulation across the whole sample without granular effects. For a thin film, $J_c$ = 15$\Delta$m/(V r), where $\Delta$m is the width of the hysteresis loop in emu, V the film volume in cm$^3$, and r the radius corresponding to the total area of the sample size, and was calculated from $\pi r^2$ = a x b (a and b are the film width and length in cm, respectively).

**Acknowledgments**:

We are grateful to John Fournelle, M. Putti, A. Xu, F. Kametani, A. Gurevich, P. Li, and V. Griffin for discussions and experimental help. Work at the University of Wisconsin was supported by funding from the Department of Energy under award number DE-FG02-06ER463, while that at the NHMFL was supported under NSF Cooperative Agreement DMR-0084173, by the State of Florida, and by AFOSR under grant FA9550-06-1-0474. NHMFL author (AY) is supported by a fellowship of the Japan Society for the Promotion of Science. All TEM work was carried out at the University of Michigan and was supported by the Department of Energy under grant DE-FG02-07ER46416.

**Figure Captions**

Figure 1.  Schematic model of Ba-122 deposition on STO(BTO) template grown on various oxide substrates. b, the Ba-122 unit cell and the alkaline earth titanium oxide (AETO) unit cells grown on rare earth (RE) perovskite oxide substrates unit cell. In this example only two unit cells of AETO and one unit cell of RE-perovskite are shown. There are two possible ways the Ba-122 and perovskite lattice bonding can occur. a, the FeAs layer in Ba-122 bonds strongly to RE-O terminated RETO as Ba-122 is deposited (the AE-O layer from AETO replaces the Ba layer in Ba-122). c, the Ba layer in Ba-122 bonds strongly to $TiO_2$ terminated RETO as B-122 is deposited (the Ba layer from Ba-122 replaces AE-O in the AETO).

Figure 2.  AFM image and RHEED pattern of the STO template grown on LSAT, and XRD patterns obtained on the Co-doped $BaFe_2As_2$ thin films. a, AFM image of 100 u.c. STO template grown on LSAT. The RHEED pattern of Ba-122 is shown in the inset. b, Out-of plane $\theta$-$2\theta$ XRD patterns of the film grown on 100 u.c. STO/LSAT and the film grown on bare LSAT. All non-identified small peaks are reflections of identified peaks due to CuK$\beta$ and CuK$\alpha$+$\beta$. c, Rocking curve and FWHM for (004) reflection of Ba-122. d, Azimuthal $\phi$ scan and $\Delta\phi$ of the off-axis 112 reflection of Ba-122. The vertical arrows above the bare LSAT pattern indicate a peak due to misoriented grains separated by high-angle grain boundaries.

Figure 3.  Cross-sectional TEM micrographs and corresponding selected area electron diffraction (SAED) patterns of films on a, bare LSAT, b, 100 u.c. STO/LSAT, and c, bare STO. They confirm that the films on 100 u.c. STO/LSAT and bare STO are epitaxial, while the film on bare LSAT is polycrystalline with misoriented grains. Line defects along the c-axis exist in the films grown on STO/LSAT and bare STO. The inset on the top right side of each image shows the corresponding SAED patterns, while the inset at the bottom left side of b shows the line defects.

Figure 4.  Resistivity and magnetic moment as a function of temperature a, $\rho(T)$ from room temperature to below $T_c$. b, Superconducting transition of all films; $T_{c,\ onset}$ and $T_{c,\ \rho=0}$ of the film on 100 u.c. STO/LSAT are as high as 22.8K and 21.5K, respectively. Inset shows superconducting transition of the film on bare STO. c, Magnetic moment as a function of temperature evaluated by warming after zero-field-cooling. A field of 2 mT was applied perpendicular to the (001) plane of the films after cooling to 4.2K. Inset shows an expanded view of the much smaller diamagnetic signal of the film grown on bare LSAT.

Figure 5. $J_c$ as a function of magnetic field and its angular dependence. a, Magnetization $J_c$ as a function of magnetic field at 4.2K with the field applied perpendicular to the (001) plane of the film. Inset shows a magneto-optical image obtained by zero-field-cooling to 6.6K, then magnetic field of 20 mT perpendicular to (001). b, Transport $J_c$ at 14K and 1, 4, and 8T as a function of the angle between the applied field and the surface of the Ba-122 films.

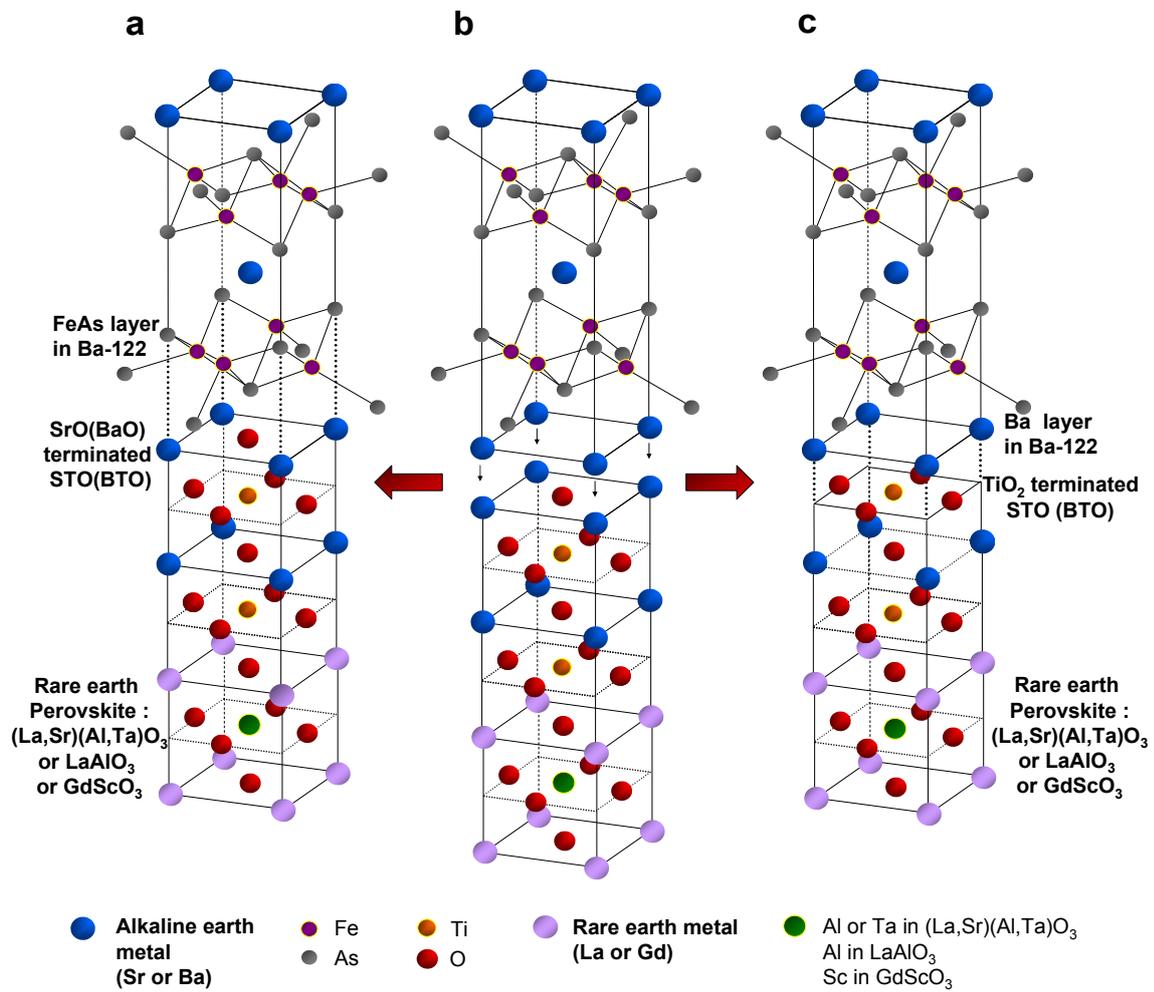

Figure 1

S. Lee *et al.*

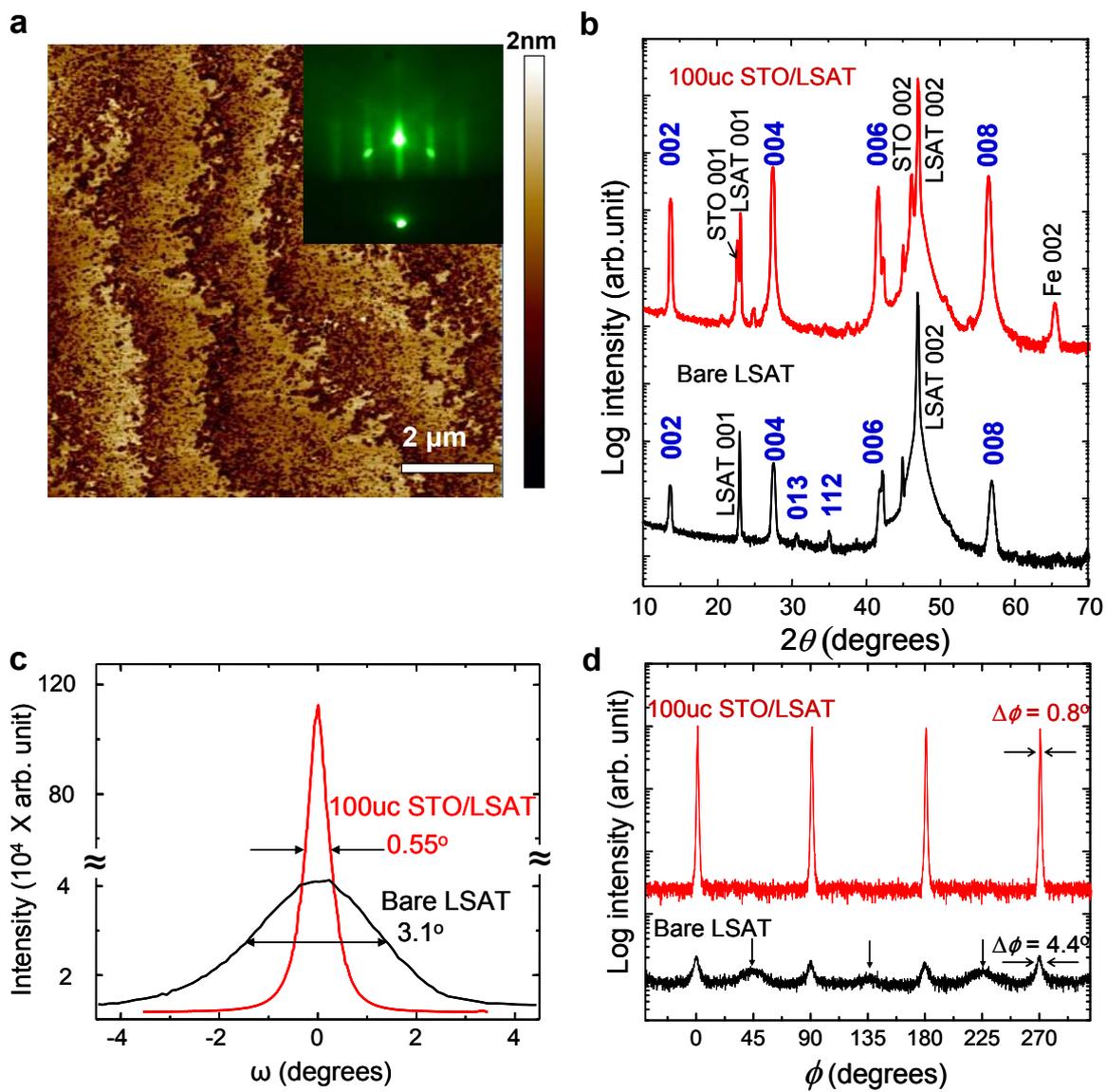

Figure 2

S. Lee *et al.*

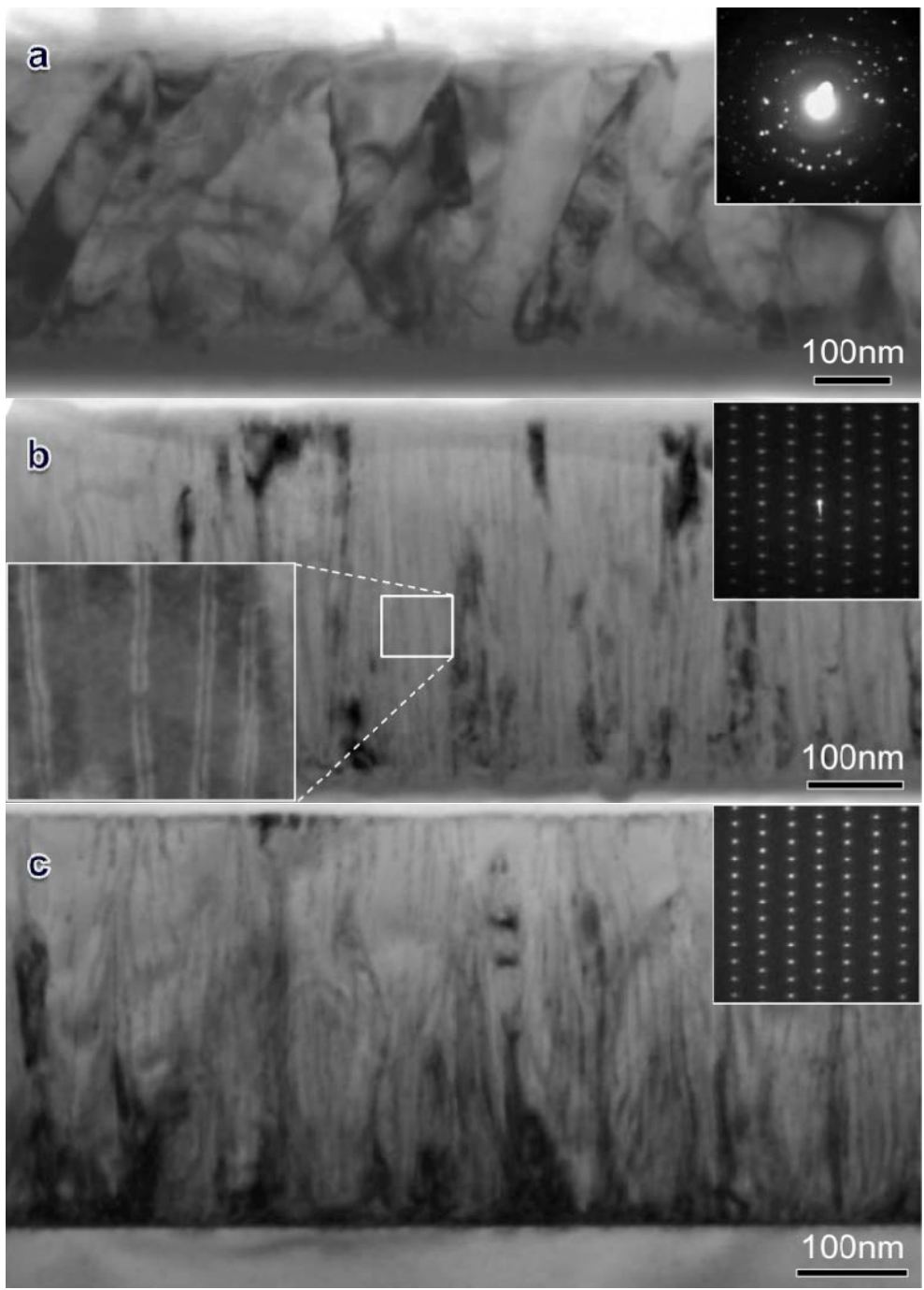

Figure 3

S. Lee *et al*.

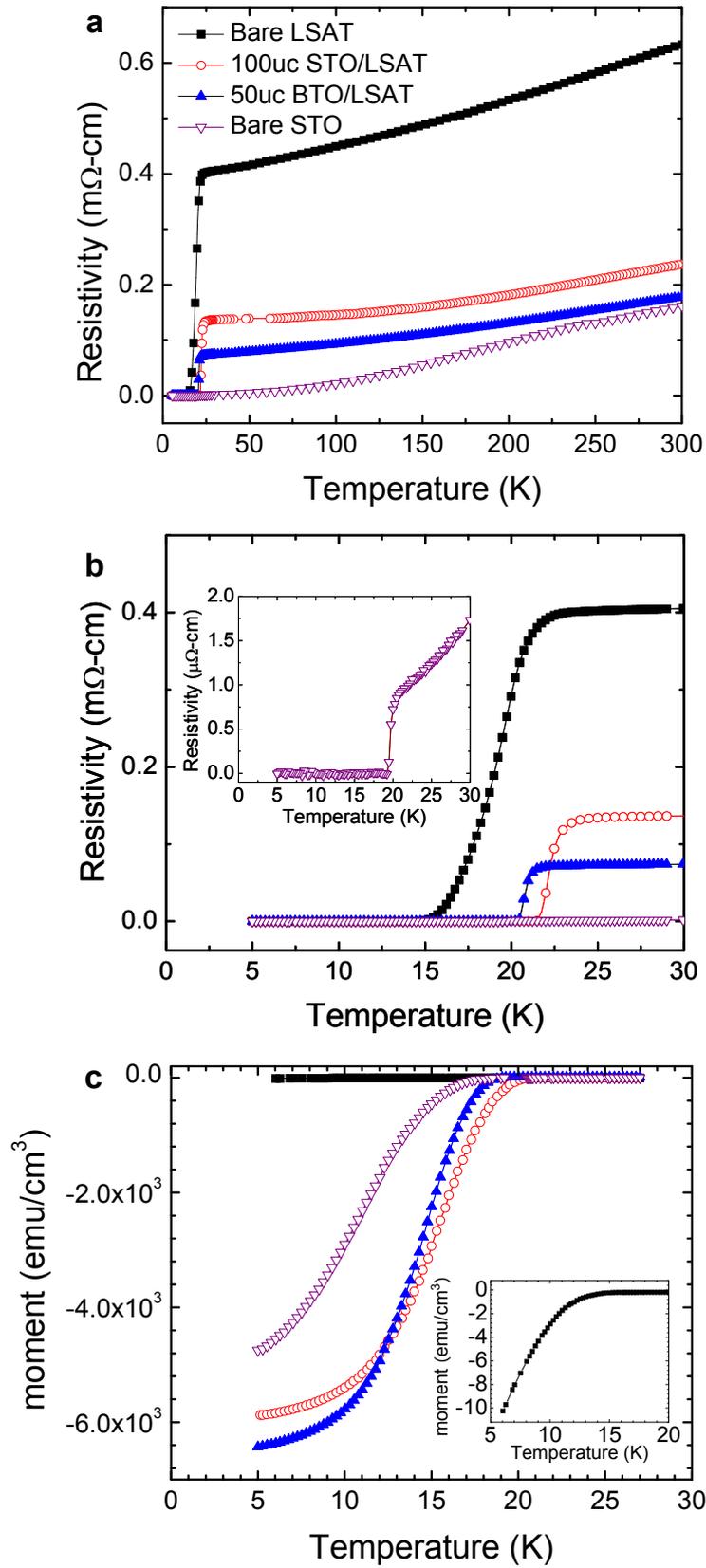

Figure 4

S. Lee *et al*.

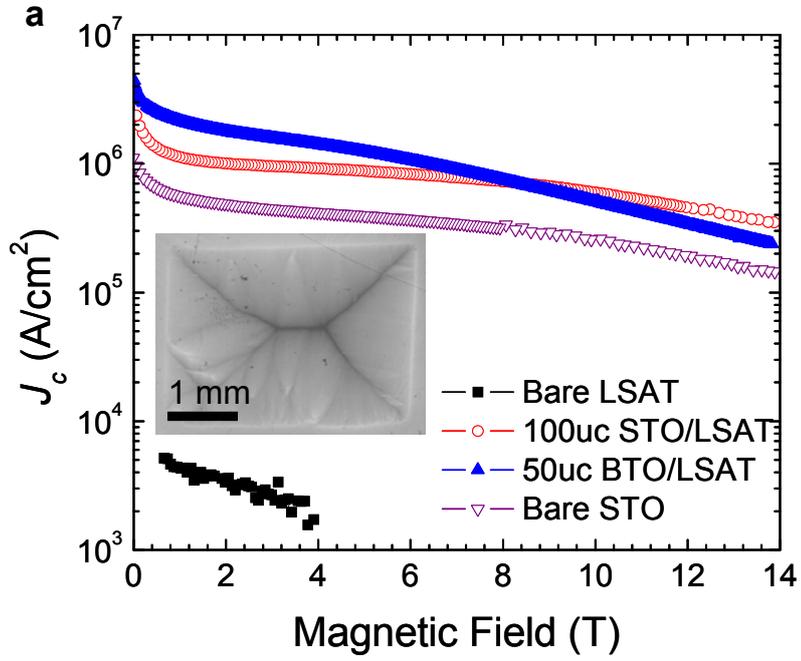

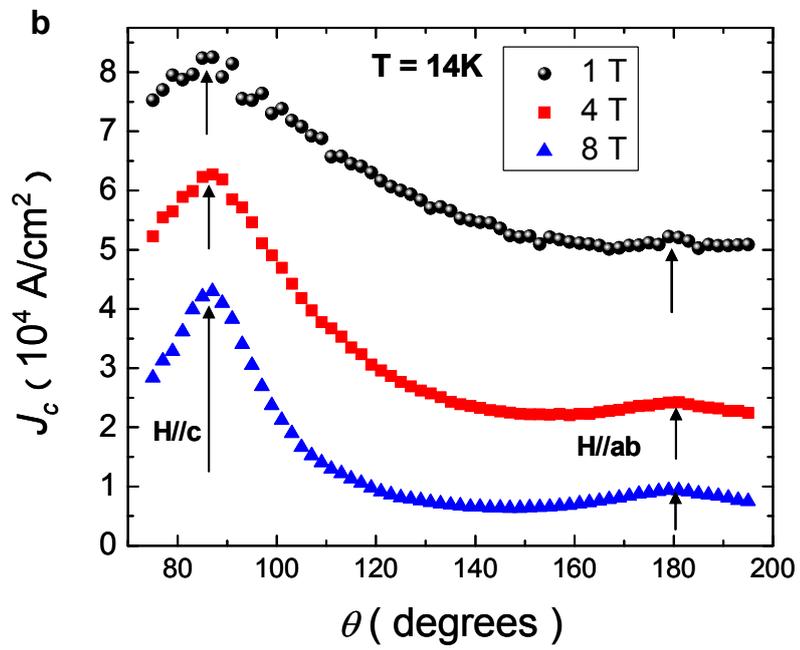

Figure 5

S. Lee *et al*.

# Template engineering of Co-doped $BaFe_2As_2$ single-crystal thin films

S. Lee[1], J. Jiang[2], C. T. Nelson[3], C. W. Bark[1], J. D. Weiss[2], C. Tarantini[2], H. W. Jang[1], C. M. Folkman[1], S. H. Baek[1], A. Polyanskii[2], D. Abraimov[2], A. Yamamoto[2], Y. Zhang[3], X. Q. Pan[3], E. E. Hellstrom[2], D. C. Larbalestier[2], C. B. Eom[1*]

[1]Department of Materials Science and Engineering, University of Wisconsin-Madison, Madison, WI 53706, USA

[2]Applied Superconductivity Center, National High Magnetic Field Laboratory, Florida State University, 2031 East Paul Dirac Drive, Tallahassee, FL 32310, USA

[3]Department of Materials Science and Engineering, The University of Michigan, Ann Arbor, Michigan 48109, USA

**Co-doped Ba-122 thin films grown on various bare and on templated substrates**

Table SI. Crystalline quality and superconducting properties of the Ba-122 films grown on various bare and on STO- or BTO-templated substrates. As a reference, bulk single-crystal properties are also shown at the bottom. Lattice mismatch with respect to the top most oxide substrate, FWHM of rocking curve of Ba-122 (004) reflection, $\Delta\phi$ of azimuthal $\phi$ scan, and existence or non-existence of extra peaks due to misoriented grains in the $\phi$ scan are shown as measures of the out-of-plane and in–plane crystalline quality. $T_{c, \rho=0}$, $\Delta T_c$, and $J_c$ characterize the superconducting properties

| Co-doped Ba122 thin films on | Lattice mismatch (%) | FWHM $\Delta\omega$ (°) | FWHM $\Delta\phi$ (°) | Extra peaks in $\phi$ scan | $T_{c,\rho=0}$ (K) | $\Delta T_c$ (K) | $J_c$ (4.2K, SF) (A/cm$^2$) |
|---|---|---|---|---|---|---|---|
| **Bare STO** | -1.7 | 1.43 | 1.6 | No | 19.2 | 0.7 | 1.1X10$^6$ |
| **Bare LSAT** | -2.6 | 3.10 | 4.4 | Yes | 14.3 | 6.8 | < 1X10$^4$ |
| **100uc STO/LSAT** | -2.6 | 0.55 | 0.8 | No | 21.5 | 1.3 | 2.9X10$^6$ |
| **50uc BTO/LSAT** | -2.6 | 0.17 | 0.6 | No | 20.8 | 1.3 | 4.5X10$^6$ |
| Bare LAO | -4.7 | 2.10 | 4.1 | Yes | 15.3 | 2.5 | |
| 50uc STO/LAO | -4.7 | 0.95 | 1.2 | No | 19.8 | 1.4 | |
| Bare GSO | ~0 | 0.62 | 0.7 | Yes | 16.5 | 4.4 | |
| 50uc STO/GSO | ~0 | 0.24 | 0.3 | No | 20.9 | 1.2 | |
| 100uc BTO/GSO | ~0 | 0.24 | 0.5 | No | 19.0 | 1.3 | |
| Bulk single crystal Ba122 | | 0.18 ref.1 | | | 21.4 ref.2 | 0.6 ref.2 | 0.4 X10$^6$ ref.3 |